\documentclass[pra,notitlepage,amsmath,amssymb]{revtex4}

\usepackage{graphicx}
\usepackage{hyperref}

\def\R{\operatorname{Re}}
\def\I{\operatorname{Im}}
\def\bbR{{\mathbb R}}
\def\bbC{{\mathbb C}}

\begin{document}

\title[Counting free fermions on a line\ldots]{Counting free fermions on a line: 
a Fisher--Hartwig asymptotic expansion 
for the Toeplitz determinant in the double-scaling limit}

\author{Dmitri A.~Ivanov}
\affiliation{
Institute for Theoretical Physics, ETH Z\"urich,
8093 Z\"urich, Switzerland}
\affiliation{
Institute for Theoretical Physics, University of Z\"urich, 
8057 Z\"urich, Switzerland}

\author{Alexander G.~Abanov}
\affiliation{Department of Physics and Astronomy,
Stony Brook University,  Stony Brook, NY 11794, USA}

\author{Vadim V.~Cheianov}
\affiliation{Physics Department, Lancaster University,
Lancaster, LA1 4YB, UK}

\begin{abstract}
We derive an asymptotic expansion for a Wiener--Hopf determinant
arising in the problem of counting one-dimensional free fermions
on a line segment at zero temperature. 
This expansion is an extension of the
result in the theory of Toeplitz and Wiener--Hopf determinants 
known as the generalized Fisher--Hartwig conjecture.
The coefficients of this expansion are conjectured to obey
certain periodicity relations, which renders the expansion
explicitly periodic in the ``counting parameter''.
We present two methods to calculate these coefficients and
verify the periodicity relations order by order:
the matrix Riemann--Hilbert problem and the
Painlev\'e V equation.
We show that the expansion coefficients are polynomials 
in the counting parameter and list explicitly first several 
coefficients.
\end{abstract}

\maketitle

\tableofcontents

\section{Introduction and motivation}
\label{sec:motivation}

Toeplitz determinants are determinants of matrices whose elements depend
only on the difference of the matrix indices:
\begin{equation}
D_N=\det_{1\le i,j \le N} a_{i-j}\, .
\label{Toeplitz-general}
\end{equation}
They occur in many topics of theoretical physics: 
statistical physics \cite{1963-MPW,2006-Basor},
random-matrix theory \cite{1993-TracyWidom,1994-Widom}, 
full counting statistics of fermionic systems \cite{2011-AIQ},
non-equilibrium bosonization \cite{2010-GutmanGefenMirlin}, 
etc. Typically, in physical applications, one is interested in 
the behaviour of a Toeplitz determinant (\ref{Toeplitz-general})
as the matrix size $N$ tends to infinity. For rapidly 
decaying matrix elements $a_{i-j}$,
the leading exponential dependence of $D_N$ on $N$ can be easily understood on 
physical grounds: the coefficient in the exponent is given by the 
average logarithm
of the ``symbol'' (the Fourier transform of $a_{i-j}$) of the Toeplitz matrix:
\begin{equation}
D_N \sim A\, \exp\left[N\oint\frac{dk}{2\pi} \ln\sigma(k) \right]\, ,
 \qquad
\sigma(k)=\sum_{m=-\infty}^{+\infty} a_m e^{-ikm}\, .
\label{Szego}
\end{equation}
This result is known as the strong Szeg\H{o} theorem 
\cite{BoettcherSilbermann-book,1952-Szego}
[the theorem also gives the coefficient $A$ in terms of $\sigma(k)$].
If the matrix elements $a_{i-j}$
decay slowly (or, equivalently, if the symbol $\sigma(k)$ has singularities),
the exponential dependence (\ref{Szego}) is complemented by power-law 
prefactors: the result known as the Fisher--Hartwig 
formula \cite{1968-FisherHartwig,1978-Basor,1991-BasorTracy,2001-Ehrhardt,%
BoettcherSilbermann-book,2010-DeiftItsKrasovsky,2010-Krasovsky}.

A particularly interesting feature of the Fisher--Hartwig formula comes from
the ambiguity in choosing the branch of the logarithm in Eq.~(\ref{Szego})
in the case of a singular $\sigma(k)$. As a result, one obtains
multiple branches of the asymptotic dependence of $D_N$ on $N$, and the
traditional Fisher--Hartwig formula prescribes selecting the leading one
(having the prefactor with the largest power of $N$).
If the Toeplitz matrix depends on a parameter, the leading branch may switch
as a function of this parameter: in this case, the asymptotic behavior of $D_N$
depends on it nonanlytically. At the switching point, two branches
are equally relevant, and the correct asymptotic behavior is given in this
case by a simple sum of these two branches. This prescription known as the
generalized Fisher--Hartwig conjecture \cite{1991-BasorTracy} 
was recently proven in Ref.~\onlinecite{2010-DeiftItsKrasovsky}.

In recent literature, it was conjectured
that subleading branches of the Fisher--Hartwig formula do not need to
be discarded, but they provide an accurate description of subleading
terms in an asymptotic expansion of $D_N$ as $N$ tends to infinity
\cite{2005-FranchiniAbanov,2011-GutmanGefenMirlin,2010-CalabreseEssler,2009-Kitanine-2}.
Moreover, each of the Fisher--Hartwig branches (\ref{Szego}) may, in turn,
be improved by including corrections as a usual power series in $1/N$.
It was further conjectured that a {\it full} asymptotic series 
for $D_N$ may be obtained as the sum of {\it all} Fisher--Hartwig branches, 
in which {\it all} terms in the $1/N$ expansion are kept 
\cite{2009-Kitanine,2008-Kozlowski}. A particular case of this conjecture
was also proposed in Ref.~\onlinecite{2011-AIQ} for the Toeplitz determinant 
describing the full counting statistics of
one-dimensional free fermions. It was verified numerically that,
in this example, the first subleading $1/N$ terms in the leading and
subleading Fisher--Hartwig branches reproduce several terms in an
asymptotic expansion of $D_N$.

In this work, we support this conjecture by an explicit calculation 
for the problem
of free fermions on a continuous line, which is a limiting
case of the lattice model considered in Ref.~\onlinecite{2011-AIQ}.
In this limit, $N$ is replaced by a continuous parameter $x$,
the Toeplitz determinant becomes a Fredholm determinant
(more specifically, a Wiener--Hopf determinant with a piecewise constant
symbol \cite{1983-BasorWidom,1994-BoettcherSilbermannWidom}), 
and we may use methods developed for the latter 
\cite{1980-JMMS,1993-TracyWidom,KorepinEtAl-book,2003-ChZ}.
We derive a complete asymptotic expansion consistent with
the above-mentioned genrealization of the Fisher--Hartwig conjecture.
A proof of this new conjecture is still missing: it amounts
to certain ``periodicity relations'' on the coefficients of this expansion.
However, we present a systematic algorithm for calculating the 
coefficients to an arbitrarily high order in $1/x$, which allows us
to verify these periodicity relations order by order.
We have verified the periodicity relations up to the 15th order in $1/x$, 
and conjecture that they hold to all orders.

\section{Formulation of the problem}
\label{sec:formulation}

We begin our definitions with a description of a relevant
physical problem of full counting statistics of free fermions on a segment
of an infinite line. Consider free fermions in one dimension at zero temperature.
Their multi-particle state is characterized by a single parameter:
the wave vector $k_F$ such that all states with wave vectors $|k|<k_F$
are filled and all states with $|k|>k_F$ are empty. We are interested in
the expectation value $\langle \exp [2\pi i\kappa \hat{Q}] \rangle$,
where $\hat{Q}$ is the operator of the number of particles on a given line
segment of length $L$ and $\kappa$ is an auxiliary 
``counting parameter'' \cite{1993-LevitovLesovik,2011-AIQ,1990-Its,note-kappa}.
This expectation value may be re-expressed as a determinant of a 
single-particle operator \cite{2002-Klich,2009-AI,2011-AIQ,KorepinEtAl-book}:
\begin{equation}
\langle \exp [2\pi i\kappa \hat{Q}] \rangle =
\det \left[1+n_F\left(e^{2\pi i \kappa Q}-1\right)\right]
= \det \left[1+n_F Q \left(e^{2\pi i \kappa}-1\right)\right]
\, .
\label{single-particle-det}
\end{equation}
Here $Q$ in the right-hand side is the single-particle projector
on the line segment in real space,
\begin{equation}
Q=\begin{cases}
1 & \text{if $0<q<L$}\, , \\
0 & \text{otherwise}
\end{cases}
\end{equation}
($q$ is the coordinate on the line) and $n_F$ is the projector
on the occupied states in the Fourier space,
\begin{equation}
n_F=\begin{cases}
1 & \text{if $|k|<k_F$}\, , \\
0 & \text{otherwise}\, .
\end{cases}
\label{nF-def}
\end{equation}
The determinant (\ref{single-particle-det}) 
is of the Wiener--Hopf type \cite{1983-BasorWidom,1994-BoettcherSilbermannWidom}. It 
may be understood as
a Fredholm determinant,
\begin{equation}
\det \left[1+n_F Q \left(e^{2\pi i \kappa}-1\right)\right] =
\sum_{l=0}^{\infty} \frac{(e^{2\pi i \kappa}-1)^l}{l!}
\int_0^L dq_1 \ldots \int_0^L dq_l 
\det_{1\le i,j \le l} \left[ n_F(q_i-q_j) \right]\, ,
\label{Fredholm-det}
\end{equation}
where $n_F(q_i-q_j)$ is the Fourier transform of $n_F$ defined
by Eq.~(\ref{nF-def}):
\begin{equation}
n_F(q)=\frac{\sin (k_F q)}{\pi q}\, .
\label{nF-q}
\end{equation}
Obviously, the determinant (\ref{Fredholm-det})
depends on $k_F$ and $L$ only via their product $k_F L$,
and therefore we may define
\begin{equation}
\chi(\kappa,x) =
\det \left[1+n_F Q \left(e^{2\pi i \kappa}-1\right)\right]\, ,
\qquad x=k_F L\, .
\label{chi-def1}
\end{equation}

Alternatively, we may define the same function $\chi(\kappa,x)$
by discretizing the space coordinate $q$ and then 
taking the continuous limit (as in Ref.~\onlinecite{2011-AIQ}).
Namely, we may first consider the Toeplitz determinant on
a lattice:
\begin{equation}
D_N(\kappa, k_F) = \det_{1\le i,j \le N}
\left[\delta_{ij} + n_F(i-j) \left(e^{2\pi i \kappa}-1\right)\right]
\end{equation}
[here $n_F(i-j)$ is defined by the same expression (\ref{nF-q}) 
with $n_F(0)=k_F/\pi$]
and then define the function $\chi(\kappa, x)$ as
the limit
\begin{equation}
\chi(\kappa,x)=\lim_{N\to\infty} D_N(\kappa, x/N)
\end{equation}
This type of definition is refered to as a 
``double-scaling limit'' in Ref.~\onlinecite{2010-Krasovsky}.

We have thus defined a function of two variables $\chi(\kappa,x)$.
At a given $x$, it is periodic in $\kappa$ with period one.
From the expansion (\ref{Fredholm-det}) it follows that $\chi(\kappa,x)$
is an entire function of $\kappa$ at any fixed value of $x$.
The main result of this paper is 
a conjecture of
an explicitly periodic asymptotic expansion for
$\chi(\kappa,x)$ at $x\to\infty$,
Eqs.\ (\ref{GFH-1}) and (\ref{GFH-2}).
As announced in the introduction,
the leading asymptotics of $\chi(\kappa,x)$ at $x\to\infty$
is discontinuous in $\kappa$ (at points $\R(\kappa)=l+1/2$
with integer $l$), and our asymptotic expansion describes
in full detail the development of this discontinuity
related to the switching of Fisher--Hartwig branches.

\section{Main results}
\label{sec:results}

In Section~\ref{sec:RH} below, we 
derive the asymptotic expansion (\ref{expansion-general})
for the function $\chi(\kappa, x)$. This form of the asymptotic
expansion is proven, provided $\R(\kappa) \ne l+1/2$ for any integer $l$,
and the coefficients $L_j(\kappa,x)$ are computable, as Laurent 
series in $1/x$, iteratively order by order.

Furthermore, we conjecture that the coefficients $L_j(\kappa,x)$
obey the ``periodicity relations'' (\ref{periodicity-conjecture}),
which brings the expansion (\ref{expansion-general}) to
an explicitly periodic form. Under this assumption
(which we verified to many orders in $1/x$), the expansion 
(\ref{expansion-general}) may be brought to the form
\begin{equation}
\chi(\kappa, x)= \sum_{j=-\infty}^{+\infty} \chi_*(\kappa+j, x)\, ,
\label{GFH-1}
\end{equation}
where
\begin{equation}
\chi_*(\kappa,x)=\exp\left[2i\kappa x - 2\kappa^2 \ln x + C(\kappa)
+\sum_{n=1}^{\infty} f_n(\kappa)\, (ix)^{-n} \right]\, .
\label{GFH-2}
\end{equation}
This asymptotic expansion agrees with the general conjecture
proposed in Refs.~\onlinecite{2009-Kitanine,2008-Kozlowski}
and with a more explicit formula conjectured
in Ref.~\onlinecite{2011-AIQ}.

The sum in Eq.~(\ref{GFH-1}) corresponds to adding together all different
Fisher--Hartwig branches, and the sum in Eq.~(\ref{GFH-2}) includes
all $(1/x)^n$ corrections within a given branch. The coefficient $C(\kappa)$
is given by 
\cite{1983-BasorWidom,1994-BoettcherSilbermannWidom,2003-ChZ,2011-AIQ}:
\begin{equation}
C(\kappa)=2\ln \left[ G(1+\kappa) G(1-\kappa) \right] - 2\kappa^2 \ln 2\, ,
\label{C-coefficient}
\end{equation}
where $G(z)$ is the Barnes $G$ function \cite{BarnesG}. 
The coefficients $f_n(\kappa)$
are polynomials in $\kappa$ with real rational coefficients, and they
are odd/even in $\kappa$ at odd/even $n$, respectively. Moreover,
the lowest power of $\kappa$ in $f_n(\kappa)$ is 3 or 4 (for $n$
odd or even, respectively) \cite{note-degree}.
The first several coefficients are \cite{note-F1}:
\begin{eqnarray}
&& f_1(\kappa)=2\kappa^3 \, , \nonumber \\
&& f_2(\kappa)=\frac{5}{2}\kappa^4 \, , \nonumber \\
&& f_3(\kappa)=\frac{11}{2}\kappa^5 + \frac{1}{6}\kappa^3 \, , \nonumber \\
&& f_4(\kappa)=\frac{63}{4}\kappa^6 + \frac{13}{8}\kappa^4 \, , 
\label{f-results} \\
&& f_5(\kappa)=\frac{527}{10}\kappa^7 + 12\kappa^5 + \frac{1}{5}\kappa^3 \, , \nonumber \\
&& f_6(\kappa)=\frac{3129}{16}\kappa^8 + \frac{1931}{24}\kappa^6 + \frac{75}{16}\kappa^4 \, , \nonumber \\
&& f_7(\kappa)=\frac{175045}{224}\kappa^9 + \frac{8263}{16}\kappa^7 + \frac{2155}{32}\kappa^5
+ \frac{45}{56} \kappa^3 \, . \nonumber
\end{eqnarray}
In Sections \ref{sec:RH} and \ref{sec:Painleve} we give algorithms for
calculating the coefficients $f_n(\kappa)$ order by order up to an
arbitrary large $n$.

In this work, we prove neither the expansion (\ref{GFH-1})--(\ref{GFH-2}) nor
even its weaker form (\ref{expansion-general}) at points $\R(\kappa) = l+1/2$. However
we conjecture that it is also valid there, so that the expansion 
(\ref{GFH-1})--(\ref{GFH-2}) is in fact a uniform asymptotic expansion on any compact
subset of $\kappa$.

\section{Derivation using the matrix Riemann--Hilbert problem}
\label{sec:RH}

In this section we show how the expansion
(\ref{GFH-1}), (\ref{GFH-2}) can be derived by means of the 
asymptotic solution of the associated Riemann--Hilbert problem. 
The relationship between the the matrix Riemann--Hilbert problem,
integrable partial differential equations and 
Fredholm determinants of integrable kernels, 
in particular the sine kernel, was exploited in different 
contexts such as classical inverse scattering problem, 
random matrix theory and quantum integrable systems 
\cite{1990-Its,1990-ItsIzergin,1992-Its,1993-Its,1997-DeiftItsZhou,%
1998-Gohmann,2000-FujiiWadati,2001-Shiroishi,2003-ChZ}.
The asymptotic solution of the matrix Riemann--Hilbert problem for 
the generalized sine kernel with Hartwig--Fisher singularities 
was developed in Refs.~\onlinecite{2003-ChZ,2009-Kitanine,2010-DeiftItsKrasovsky}.
In this section we follow the notations of Ref.~\onlinecite{2003-ChZ}. 
In that work, the logarithmic derivative of the function 
$\chi(\kappa,x)$ was expressed in terms
of a solution of a certain Riemann--Hilbert matrix problem, and then
this problem was solved to the leading order in $x$. In this section,
we solve the same Riemann--Hilbert matrix problem in terms of a
series in $1/x$ whose coefficients may be calculated iteratively, 
order by order.

We do not repeat here the full argument of Ref.~\onlinecite{2003-ChZ}, 
but start with formulating their result relevant for our calculation. 
We refer the reader to the original paper for its derivation. 
The logarithmic derivative of $\chi(\kappa,x)$ was expressed there as
\begin{equation}
\frac{\partial}{\partial x} \ln \chi(\kappa,x)=
2 i\kappa - i \lim_{k \to \infty} k\left[S_\infty(k)_{11}-1\right]\, ,
\label{log-derivative}
\end{equation}
where $S_\infty (k)_{11}$ is the upper-left-corner matrix element of
the $2\times 2$ matrix $S_\infty (k)$, which solves the Riemann--Hilbert
problem (in the complex variable $k$) formulated below.

\begin{figure}
	\includegraphics[width=0.4\textwidth]{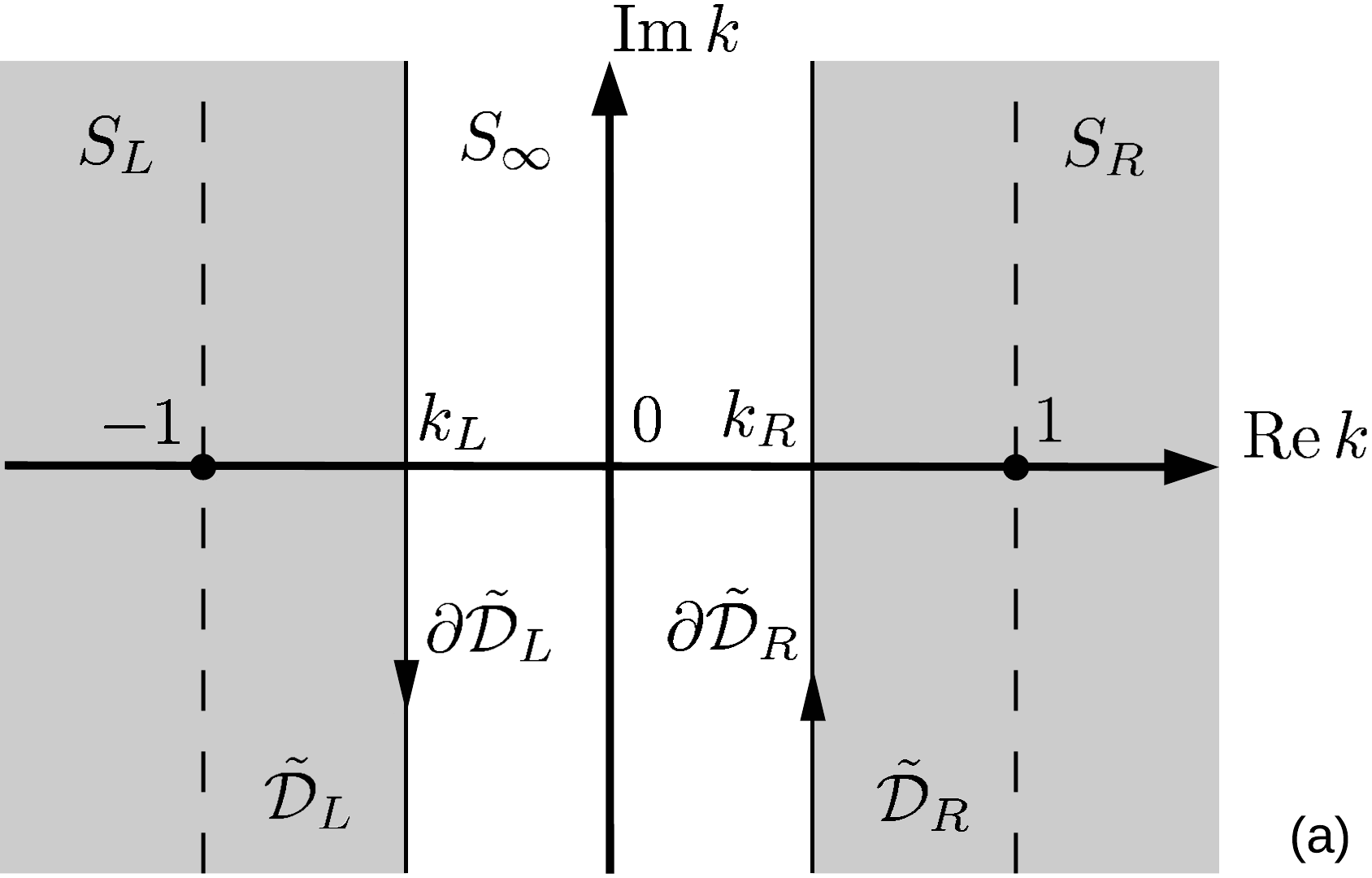}\hskip 0.1\textwidth
	\includegraphics[width=0.4\textwidth]{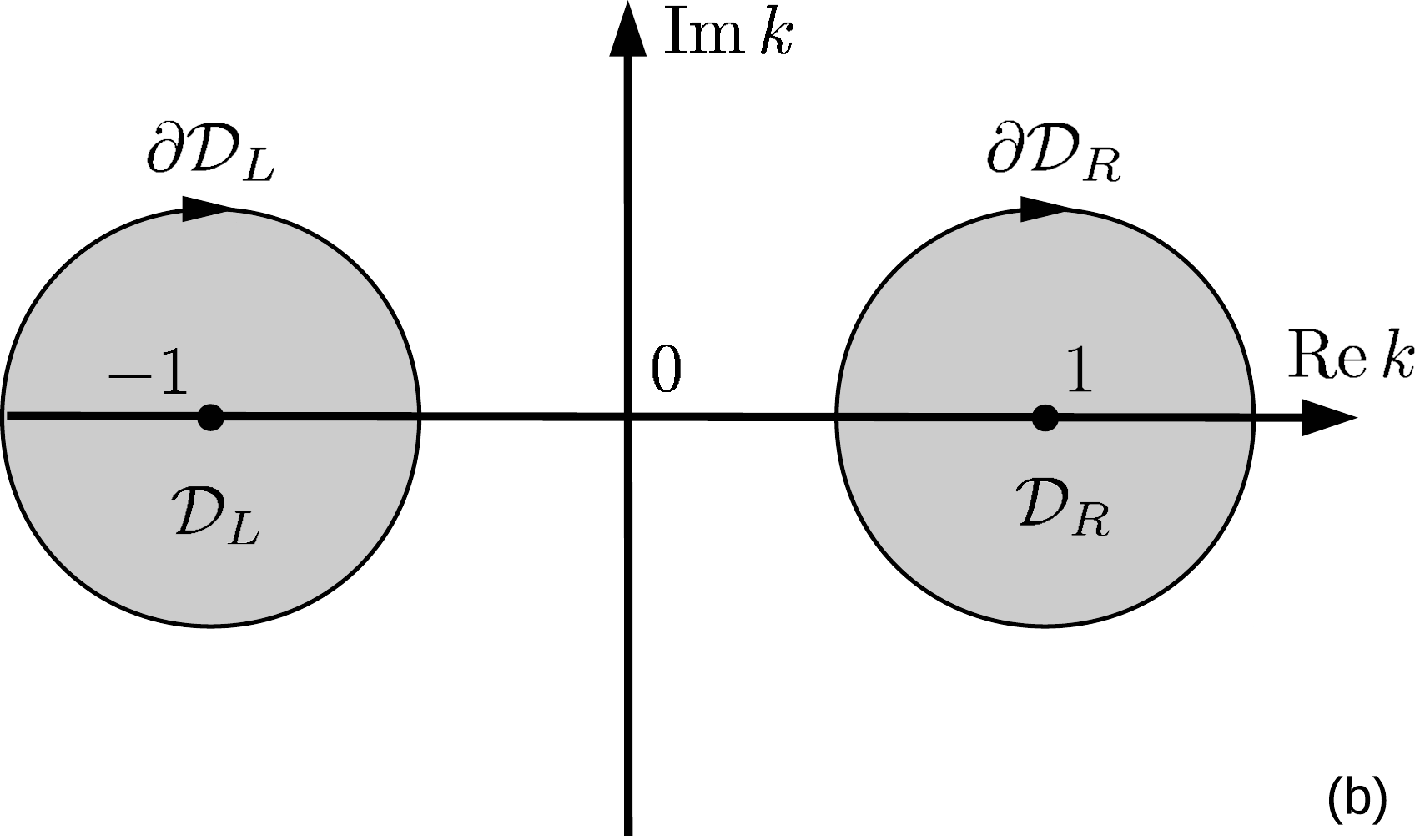}
\caption{{\bf (a)} Initial choice of domains for the matrix
Riemann--Hilbert problem (\ref{RH-problem}). The dashed lines
denote the branch cuts of $\theta_R(k)$ in $\tilde{\cal D}_R$
and of $\theta_L(k)$ in $\tilde{\cal D}_L$. The arrows at the
boundaries $\partial \tilde{\cal D}_{R,L}$ denote their orientation
in the integral (\ref{S-solution}).
{\bf (b)} The choice of domains used in Ref.~\onlinecite{2003-ChZ} and
in our formula (\ref{I-general}).}
\label{fig:domains}
\end{figure}

Define regions $\tilde{\cal D}_R$ and $\tilde{\cal D}_L$ as $\R k>k_R$ and
$\R k < k_L$, respectively, where the region boundaries are chosen as
$-1<k_L<k_R<1$ (see Fig.~\ref{fig:domains}a). This choice of regions differs slightly from
Ref.~\onlinecite{2003-ChZ}, where the regions were chosen as discs around $k=\pm 1$: this
difference does not change anything in our calculation, but
simplifies  the discussion of branch cuts. In these regions, 
we define the two matrix-valued functions:
\begin{equation}
\theta_R(k)=
\left( \begin{array}{cc}
\tilde{\Psi} [-\kappa, -ix(k-1)] &
\displaystyle\frac{a_R [x(k+1)]^{-2\kappa}}{x(k-1)} 
\tilde{\Psi} [1+\kappa, ix(k-1)] \\
\displaystyle\frac{b_R [x(k+1)]^{2\kappa}}{x(k-1)}  
\tilde{\Psi} [1-\kappa,-ix(k-1)] &
\tilde{\Psi} [\kappa, ix(k-1)] 
\end{array} \right)
\end{equation}
in the region $\tilde{\cal D}_R$ and
\begin{equation}
\theta_L(k)=
\left(\begin{array}{cc}
\tilde{\Psi} [\kappa, ix(k+1)] &
\displaystyle\frac{a_L [x(k-1)]^{2\kappa}}{x(k+1)} 
\tilde{\Psi} [1-\kappa, ix(k+1)] \\
\displaystyle\frac{b_L [x(k-1)]^{-2\kappa}}{x(k+1)}  
\tilde{\Psi} [1+\kappa,-ix(k+1)] &
\tilde{\Psi} [-\kappa, -ix(k+1)] 
\end{array} \right)
\end{equation}
in the region $\tilde{\cal D}_L$. Here the constants $a_R$ and $b_R$
are the same as in Ref.~\onlinecite{2003-ChZ},
\begin{equation}
a_R= -\frac{i\pi e^{i\pi\kappa}e^{ix}}{\Gamma^2(-\kappa)}\, , \qquad
b_R= -\frac{i\pi e^{-i\pi\kappa}e^{-ix}}{\Gamma^2(\kappa) 
\sin^2 (\pi\kappa)}\, ,
\end{equation}
the coefficients $a_L$ and $b_L$ are defined as 
\begin{equation}
b_L= -\frac{a_R}{\sin^2 (\pi\kappa)}\, , \qquad
a_L= - b_R \sin^2 (\pi\kappa)\, ,
\end{equation}
and we have introduced a shorthand notation
\begin{equation}
\tilde\Psi(a,w)=w^a \Psi(a,1;w)\, ,
\end{equation}
where $\Psi(a,1;w)$ is the Tricomi function \cite{1953-BatemanErdelyi}. 
The functions $\Psi(a,1;w)$ and $\tilde\Psi(a,w)$ are defined to
have branch-cut discontinuities along the negative real axis $w \in {\bbR}_-$.
Then the matrices $\theta_R(k)$ and $\theta_L(k)$ have discontinuities
along the lines $\R k =\pm 1$, respectively. Note however that
those discontinuities decay exponentially in $|x(k\mp 1)|$, so that
$\theta_R(k)$ and $\theta_L(k)$ have regular asymptotic expansions
in inverse powers of $x(k\mp 1)$ (see Ref.~\onlinecite{2003-ChZ} 
for more detail). 
We now formulate the Riemann--Hilbert problem for a matrix $S(k)$:
\begin{eqnarray}
{\rm (i)~~} &&  S(k) {\rm ~~is~analytic~in~~} {\bbC} \backslash
(\partial\tilde{\cal D}_R \cup \partial\tilde{\cal D}_L) \, ;
\nonumber \\
{\rm (ii)~~} && S_\infty(k)=S_{R,L}(k)\theta_{R,L} {\rm ~at~} 
k\in \partial\tilde{\cal D}_{R,L}\, ; 
\label{RH-problem} \\
{\rm (iii)~~} && S_\infty(k) \to I {\rm ~~as~~} k\to\infty\, , 
\nonumber
\end{eqnarray}
where we denote by $S_{R,L,\infty}$ the restrictions of $S(k)$ onto
$\tilde{\cal D}_R$,  $\tilde{\cal D}_L$, and 
${\bbC} \backslash(\tilde{\cal D}_R \cup \tilde{\cal D}_L)$,
respectively. The matrix $S_\infty(k)$ obtained as a solution to this
Riemann--Hilbert problem can be then used to find the logarithmic
derivative of $\chi(\kappa,x)$, according to Eq.~(\ref{log-derivative}).
This relation is exact and is a minor reformulation of
results obtained in Sections 3 and 4 of Ref.~\onlinecite{2003-ChZ}.

We can now calculate the asymptotic expansion for $S_\infty(k)$ by
expanding $\theta_{R,L}$ in powers of $1/x$ and matching the matrix
at the contours $\partial\tilde{\cal D}_{R,L}$ order by order. In this
expansion, we will treat all powers of $x^\kappa$ and of $e^{ix}$ as
terms of order {\it zero} with respect to $x$: in other words, we
will collect together all terms with the same integer powers of $x$,
while letting the coefficients to depend on $x^\kappa$ and $e^{ix}$.
As we shall see below, such a method indeed produces an asymptotic expansion
for $S_\infty(k)$ within the interval $|\R(\kappa)|<1/2$.

We start with the (asymptotic) expansion
\begin{equation}
\tilde\Psi(a,w)=\sum_{n=0}^\infty \frac{p_n(a)}{w^n}\, ,
\qquad {\rm where} \qquad
p_n(a)=\frac{(-1)^n}{n!} a^2 (a+1)^2 \ldots (a+n-1)^2
\label{Tricomi-expansion}
\end{equation}
to expand
\begin{equation}
\theta_{R,L}= I + \frac{1}{x} \theta_{R,L}^{(1)} + \frac{1}{x^2} \theta_{R,L}^{(2)} + \ldots\, ,
\end{equation}
where
\begin{equation}
\theta_R^{(n)} = \frac{1}{(k-1)^n}
\left( \begin{array}{cc}
i^n p_n(-\kappa) & 
    (-i)^{n-1} a_R p_{n-1}(1+\kappa)
    \left[x(k+1)\right]^{-2\kappa} \\
    i^{n-1} b_R p_{n-1}(1-\kappa)
    \left[x(k+1)\right]^{2\kappa} &
(-i)^n p_n(\kappa) 
\end{array}\right)
\label{theta-expansion-R}
\end{equation}
and
\begin{equation}
\theta_L^{(n)} = \frac{1}{(k+1)^n}
\left( \begin{array}{cc}
i^n p_n(\kappa) & 
    (-i)^{n-1} a_L p_{n-1}(1-\kappa)
    \left[x(k-1)\right]^{2\kappa} \\
    i^{n-1} b_L p_{n-1}(1+\kappa)
    \left[x(k-1)\right]^{-2\kappa} &
(-i)^n p_n(-\kappa) 
\end{array}\right) 
\label{theta-expansion-L}
\end{equation}
(note that $\theta_{L,R}^{(n)}$ depend on $x$ themselves, but only
``weakly'', via $x^{\pm2\kappa}$).

Now the Riemann--Hilbert problem (\ref{RH-problem}) may be solved
iteratively (order by order) in terms of an expansion
\begin{equation}
S(k)=I+\frac{1}{x} S^{(1)}(k) + \frac{1}{x^2} S^{(2)}(k) + \ldots\, ,
\label{S-series}
\end{equation}
where $S^{(n)}(k)$ are polynomials in $x^{\pm2\kappa}$.
Note that this method slightly
differs from the approach used in Ref.~\onlinecite{2003-ChZ}, 
where the ansatz
(3.53) allowed to partly resum the series (\ref{S-series}).
We denote the functions $S^{(n)}$ in the three domains 
$\tilde{\cal D}_R$,  $\tilde{\cal D}_L$, and 
${\bbC} \backslash(\tilde{\cal D}_R \cup \tilde{\cal D}_L)$
by $S_R^{(n)}$, $S_L^{(n)}$, and $S_\infty^{(n)}$, respectively.
To the first order, we find
\begin{equation}
S_\infty^{(1)}(k)=S_{R,L}^{(1)}(k) + \theta_{R,L}^{(1)}(k)
\quad {\rm at} \quad
k\in \partial\tilde{\cal D}_{R,L}\, ,
\label{S-1-equation}
\end{equation}
while the general equation at the $n$-th order is
\begin{equation}
S^{(n)}_\infty(k)=S^{(n)}_{R,L}(k) + S^{(n-1)}_{R,L}(k) \theta^{(1)}_{R,L}(k)
+ \ldots + S^{(1)}_{R,L}(k) \theta^{(n-1)}_{R,L}(k) + \theta^{(n)}_{R,L}(k)
\quad {\rm at} \quad
k\in \partial\tilde{\cal D}_{R,L}\, .
\label{S-n-equation}
\end{equation}
Using the analyticity of $S_R^{(n)}$, $S_L^{(n)}$, and $S_\infty^{(n)}$
in the domains $\tilde{\cal D}_R$,  $\tilde{\cal D}_L$, and 
${\bbC} \backslash(\tilde{\cal D}_R \cup \tilde{\cal D}_L)$, respectively,
and the boundary condition $S^{(n)}_\infty(k) \to 0$ at $k \to \infty$,
we can solve these equations by the Cauchy integral formula. At the first
order, solving Eq.~(\ref{S-1-equation}), we find
\begin{equation}
S^{(1)}(k) = 
\oint_{\partial\tilde{\cal D}_R} \frac{dk'}{2\pi i}
\frac{1}{k'-k} \theta^{(1)}_R(k') + 
\oint_{\partial\tilde{\cal D}_L} \frac{dk'}{2\pi i}
\frac{1}{k'-k} \theta^{(1)}_L(k')
\label{S-1-solution}
\end{equation}
and, more generally, at the $n$-th order, the solution to
Eq.~(\ref{S-n-equation}) reads
\begin{equation}
S^{(n)}(k) = \sum_{\alpha=R,L} \oint_{\partial\tilde{\cal D}_\alpha} \frac{dk'}{2\pi i}
\frac{1}{k'-k}  \left[ S^{(n-1)}_{\alpha}(k') \theta^{(1)}_{\alpha}(k')
+ \ldots + S^{(1)}_{\alpha}(k') \theta^{(n-1)}_{\alpha}(k') + 
\theta^{(n)}_{\alpha}(k') \right]\, .
\label{S-solution}
\end{equation}
These formulas produce the components $S_\infty^{(n)}(k)$, $S_R^{(n)}(k)$,
and $S_L^{(n)}(k)$ depending on the location of the point $k$. 
At this stage of the calculation, it is technically convenient
to deform the integration contours $\partial\tilde{\cal D}_\alpha$
into the boudaries $\partial{\cal D}_\alpha$ of some nonoverlapping
discs ${\cal D}_R$ and ${\cal D}_L$ centered at $k=1$ and $k=-1$,
respectively, as in Ref.~\onlinecite{2003-ChZ} (See Fig.~\ref{fig:domains}b). 
This deformation is allowed, since the ingtegrations converge rapidly at infinity 
and the cuts of the matrix functions $\theta_{R,L}(k)$ disappear from the asymptotic 
expansion.

The formula (\ref{S-solution}), in principle, solves our problem:
knowing the expansions (\ref{theta-expansion-R}) and (\ref{theta-expansion-L}),
we iteratively calculate $S^{(n)}(k)$ from (\ref{S-solution}) and then 
extract the logarithmic derivative of $\chi(\kappa,x)$ using (\ref{log-derivative}).
As a result, we obtain the asymptotic series \cite{note-expansion}
\begin{equation}
\frac{\partial}{\partial x}\ln\chi(\kappa,x)
= 2i\kappa + \sum_{n=1}^\infty I_n(\kappa,x)\, ,
\label{series-1}
\end{equation}
where
\begin{equation}
I_n(\kappa,x)= 
\sum_m R_{n,m}(\kappa) \, 
x^{-n-4m\kappa} e^{2imx}
\label{series-2}
\end{equation}
is the contribution from $S_\infty^{(n)}(k)$.
The form (\ref{series-2}) of the term $I_n(\kappa,x)$ follows from
examining the explicit expression
\begin{equation}
I_n (\kappa,x) = i 
\sum_{\substack{
\{n_j\} \\  n_j \ge 1 \\ \sum_{j=1}^{s} n_j=n }}
\sum_{\substack{
\{\alpha_j\} \\ \alpha_j=R,L}}
\oint_{\partial D_{\alpha_1}^{[1]}} \frac{dk_1}{2\pi i} \ldots 
\oint_{\partial D_{\alpha_s}^{[s]}} 
\frac{dk_s}{2\pi i}
\frac{1}{k_1-k_2} \cdots \frac{1}{k_{s-1}-k_s} 
\left[\theta_{\alpha_1}^{(n_1)}(k_1) \ldots \theta_{\alpha_s}^{(n_s)}(k_s)
\right]_{11}\, ,
\label{I-general}
\end{equation}
which is obtained by an iterative application of Eq.~(\ref{S-solution}).
Here the sum is taken over all integer partitions
of $n$ into the sum of $n_j$,
and, for each such partition, over choices of the left and right
integration contour for each $j$. Furthermore, the integration contours
are ordered in such a way that the contour $\partial D_{\alpha}^{[j]}$ lies
{\it inside} $\partial D_{\alpha}^{[j']}$, if $j>j'$. Every integral in 
Eq.~(\ref{I-general}) can be easily calculated by residues. The coefficients
$R_{n,m}(\kappa)$ can be easily extracted from this expression
by selecting terms with a particular power of $x$. The following properties
of these coefficients can be proven:
\begin{enumerate}
\item 
The coefficients $R_{n,m}(\kappa)$ vanish for $|m| > [n/2]$.
This means that the sum over $m$ in Eq.~(\ref{series-2}) extends only 
from $-[n/2]$ to $+[n/2]$.
\item
The coefficients $R_{n,m}(\kappa)$ have a definite parity:
$R_{n,m}(-\kappa)= (-1)^{n-1} R_{n,-m}(\kappa)$.
This follows from the symmetry $k \mapsto -k$ (see a detailed discussion
of this symmetry in Section 3 of Ref.~\onlinecite{2003-ChZ}).
\item
The coefficients $R_{n,0}(\kappa)$ are polynomials in $\kappa$
with real rational coefficients multiplied by $(-i)^{n+1}$. The smallest
possible degree of $\kappa$ contained in this polynomial is 3 or 4, 
depending on the parity of $n$. This can be easily seen for every term 
in the sum (\ref{I-general}) by using the identity 
$a_R b_R= a_L b_L = -\kappa^2$.
\item
The first coefficient in the expansion (\ref{series-1}) is 
$R_{1,0}(\kappa)= -2\kappa^2$ [calculated
directly using the formula (\ref{I-general})].
\end{enumerate}

Note that the powers of $x$ in the formal series (\ref{series-1})--(\ref{series-2})
decay if and only if $|\R(\kappa)|<1/2$. Therefore our calculation produces
an asymptotic expansion for $(\partial/\partial x )\ln\chi(\kappa,x)$ only
within this interval of values of $\kappa$.

To obtain the expansion of the function $\chi(\kappa,x)$, we 
integrate and then exponentiate the expansion (\ref{series-1}), 
(\ref{series-2}) as a formal series in $x^{-1}$. In the
resulting series, we collect together terms with the same
oscillatory prefactor $e^{2ijx}$. The result may be
further written as
\begin{equation}
\chi(\kappa,x)=\sum_{j=-\infty}^{+\infty} \exp \left[
2i(\kappa+j)x - 2 (\kappa+j)^2 \ln x \right] L_j(\kappa,x)\, ,
\label{expansion-general}
\end{equation}
where $L_j(\kappa,x)$ are Laurent series in $x$ with
coefficients depending on $\kappa$. These coefficients
may be expressed in terms of $R_{n,m}$ and vice versa,
modulo an overall numerical prefactor in all $L_j(\kappa,x)$,
which is left undetermined [it corresponds to the 
integration constant of Eq.~(\ref{series-1})]. 

It seems very plausible [and it was conjectured both
in the Toeplitz (chain) and Wiener--Hopf (continuous) cases 
\cite{2009-Kitanine,2008-Kozlowski,2011-AIQ}] that
the expansion of the form (\ref{expansion-general})
is {\em explicitly periodic} in $\kappa$, namely
\begin{equation}
L_j(\kappa,x) = L_0(\kappa+j,x)\, .
\label{periodicity-conjecture}
\end{equation}
We do not have a proof of this conjecture at the moment,
but we have verified it analytically up to the order
$x^{-15}$ in Eq.~(\ref{expansion-general}) using
the technique based on the Painllev\'e V equation, see
Section~\ref{sec:Painleve}. We conjecture
that the relations (\ref{periodicity-conjecture})
hold to all orders.

Finally, we observe that $L_0(\kappa,x)$ does not
contain negative powers of $x$ 
[and, under the periodicity conjecture 
(\ref{periodicity-conjecture}), neither do any
of $L_j(\kappa,x)$]
and, therefore may be formally written as
\begin{equation}
L_0(\kappa,x) = \exp \left[ C(\kappa) + 
\sum_{n=1}^\infty f_n(\kappa)(ix)^{-n} \right]
\end{equation}
(we included the factors $i^{-n}$ in the expansion to make
the coefficients real). 
Under the ``periodicity conjecture'' (\ref{periodicity-conjecture}),
this immediately leads to the perioduc form of the expansion
(\ref{GFH-1})--(\ref{GFH-2}).

The coefficient $C(\kappa)$ cannot be calculated by the method
described, but it is known from other 
approaches 
\cite{1983-BasorWidom,1994-BoettcherSilbermannWidom,2003-ChZ,2011-AIQ} 
and is given by Eq.~(\ref{C-coefficient}). To calculate the coefficients
$f_n(\kappa)$, we relate them to $R_{n,m}(\kappa)$.
In fact, it is sufficient to consider only the coefficients
$R_{n,0}(\kappa)$. By comparing the expansion 
(\ref{series-1}), (\ref{series-2}) wtih (\ref{GFH-1}) -- (\ref{C-coefficient}),
one finds:
\begin{eqnarray}
&& R_{2,0}(\kappa)= - i^{-1} f_1(\kappa) \, , \nonumber \\
&& R_{3,0}(\kappa)= - 2 i^{-2} f_2(\kappa)\, , \nonumber \\
&& R_{4,0}(\kappa)= - 3 i^{-3} f_3(\kappa)\, , \nonumber \\
&& R_{5,0}(\kappa)= - 4 i^{-4} \left[ f_4(\kappa) - e^{\Delta_0} \right] \, , 
\label{C-f-relation}\\
&& R_{6,0}(\kappa)= - 5 i^{-5} \left[ f_5(\kappa) - e^{\Delta_0} \Delta_1 \right] \, , \nonumber \\
&& R_{7,0}(\kappa)= - 6 i^{-6} \left[ f_6(\kappa) - e^{\Delta_0} 
\left( \Delta_2 + \frac{\Delta_1^2}{2} \right) \right]\, , \nonumber \\
&& R_{8,0}(\kappa)= - 7 i^{-7} \left[ f_7(\kappa) - e^{\Delta_0}
\left( \Delta_3 + \Delta_2 \Delta_1 + \frac{\Delta_1^3}{6} \right) \right] \, , \nonumber
\end{eqnarray}
where we denote
\begin{equation}
e^{\Delta_0} = \exp \left[ C(\kappa+1) + C(\kappa-1) - 2 C(\kappa) \right] = \frac{\kappa^4}{16}
\end{equation}
and
\begin{equation}
\Delta_{n\ge 1} = f_n(\kappa+1) + f_n(\kappa-1) -2 f_n(\kappa)\, .
\end{equation}
Note that $R_{n,0}(\kappa)$ with $n>4$ contain
cross terms (containing $\Delta_n$) arising
from Fisher--Hartwig branches with $j\ne 0$
in Eq.~(\ref{GFH-1}).
By explicitly calculating $R_{n,0}(\kappa)$ 
from Eq.~(\ref{I-general}) [we used a computer program
to perform this calculation] and solving Eqs.~(\ref{C-f-relation}),
we arrive at the results (\ref{f-results}). We also remark
that, for large $n$, the use of the formula (\ref{I-general})
is not practical for explicit calculations (the number of 
terms grows very rapidly with $n$), and Eq.~(\ref{S-solution}) seems
to be more efficient. 

We can now prove the properties of the coefficients $f_n(\kappa)$
declared in Section~\ref{sec:results}.
\begin{itemize}
\item
The reality of $f_n(\kappa)$ follows from the property 3
of $R_{n,0}(\kappa)$: indeed, if we choose $(ix)$ as the
expansion variable, then all the coefficients of the expansions become real.
\item
The parity of the coefficients $f_n(\kappa)$ follows from the
parity of $R_{n,0}(\kappa)$ (property 2). 
One can also formulate this property as an invariance of the 
whole asymptotic series with respect to the simultaneous formal
sign change of $x$ and $\kappa$ in their integer powers, while
transforming fractional powers of $x$ as $x^{4\kappa} \mapsto x^{-4\kappa}$.
\item
One can also easily verify that the cross terms in relations (\ref{C-f-relation}),
for any $R_{n,0}$, are always polynomials divisible by $\kappa^4$.
Therefore the property that $R_{n,0}$ is always divisible by $\kappa^3$
also holds for $f_n(\kappa)$.
\end{itemize}

\section{Calculation using the Painlev\'e V equation}
\label{sec:Painleve}

An alternative way to obtain the asymptotic
expansion (\ref{GFH-1}), (\ref{GFH-2}) is the use of the
Painlev\'e V equation. It was discovered in the seminal 
paper \cite{1980-JMMS} (with a simpler version of the
derivation presented later in Ref.~\onlinecite{1993-TracyWidom})
that the Fredholm determinant
(\ref{chi-def1}) considered as a function of $x$ satisfies 
an ordinary differential equation: the Painlev\'e V equation
in the Jimbo--Miwa form,
\begin{equation}
(x\sigma'')^{2} +4 (x\sigma'-\sigma)\left( x\sigma'-\sigma + (\sigma')^{2} \right) = 0\, .
\label{JM-Painleve}
\end{equation}
Here prime means the derivative with respect to $x$ and
\begin{equation}
\sigma(\kappa,x)=x\frac{\partial}{\partial x}\ln\chi(\kappa,x)\, .
\label{sigma-def}
\end{equation}
Remarkably the parameter $\kappa$ does not enter the equation 
(\ref{JM-Painleve}) itself but defines its solution
through the boundary condition:
\begin{equation}
\sigma(\kappa,x) = \frac{e^{2\pi i \kappa}-1}{\pi} x 
-\left(\frac{e^{2\pi i \kappa}-1}{\pi}\right)^{2}x^{2}
+ O(x^3) \qquad \mbox{as }x\to 0 
 \label{xto0}
\end{equation} 
[the same expansion can also be obtained from Eq.~(\ref{Fredholm-det})].
The problem is now to find the asymptotic expansion of the solution of
(\ref{JM-Painleve}) as $x\to \infty$ if the asymptotics at $x\to 0$
are given by (\ref{xto0}). This problem was addressed in 
Ref.~\onlinecite{1986-McCoyTang} who argued that the large-$x$
asymptotics may be found in the form 
(\ref{series-1}), (\ref{series-2}) \cite{note-McCoy-notation}.

It is then straightforward to calculate the coefficients
$R_{n,m}$ of the expansion (\ref{series-1}), (\ref{series-2})
order by order, by substituting this expansion into the Painlev\'e V
equation (\ref{JM-Painleve}) and starting from the first two 
known terms of the expansion \cite{note-McCoy-typo}. 
At each order, a second-order differential equation is solved
for $I_n(\kappa,x)$, with the integration constants fixed by
requiring that no extra terms are generated beyond those in
Eqs.~(\ref{series-1}), (\ref{series-2}).
After that, the derivation fully repeats
that in Section~\ref{sec:RH}: we can restore the coefficients
$f_n(\kappa)$ in our ``periodic'' form of the expansion 
(\ref{GFH-1}), (\ref{GFH-2}) from the calculated coefficients
$R_{n,m}$ by using relations (\ref{C-f-relation}).
Of course, this method produces the same asymptotic
expansion as the method of Section~\ref{sec:RH}. We have
checked it by an explicit calculation of the coefficients
$f_n(\kappa)$ up to $n=7$ and found that the two methods 
give identical results (\ref{f-results}).
Furthermore, we extended the calculation via the Painlev\'e V
equation up to $n=15$: this allowed us to calculate the
first 15 coeffiencts $f_n(\kappa)$ and verify the
expansion (\ref{GFH-1})--(\ref{GFH-2}) up to $x^{-15}$.

\section{Summary and discussion}
\label{sec:summary}

The main result of this work is the conjecture of the
``periodic form'' of the asymptotic expansion (\ref{GFH-1})--(\ref{GFH-2}) 
for the Wiener--Hopf determinant (\ref{chi-def1}). This expansion
is based on the proven form (\ref{expansion-general}), together
with the ``periodicity conjecture'' (\ref{periodicity-conjecture}).
The latter can be verified to any order in $1/x$ by an explicit calculation.

Our methods only allow to establish the asymptotic expansion
away from the Fisher--Hartwig switching points $\R(\kappa) = l+1/2$.
However, we believe that it also holds there [this is supported by
the numerical calculation of Ref.~\onlinecite{2011-AIQ} in the lattice case], 
which would result in a uniform [with respect to $\R(\kappa)$] estimate:
\begin{equation}
\chi(\kappa, x)= \sum_{j=-j_{\rm max}}^{j_{\rm max}}
\exp\left[2i(\kappa+j) x - 2(\kappa+j)^2 \ln x + C(\kappa+j)
+\sum_{n=1}^{n_0} f_n(\kappa+j)\, (ix)^{-n} \right] + o( e^{-2\I \kappa\, x} x^{-n_0})\, ,
\label{GFH-uniform}
\end{equation}
where
\begin{equation}
j_{\rm max}=\left\lfloor \sqrt{\frac{n_0}{2}}+\frac{1}{2} \right\rfloor\, ,
\end{equation}
$\lfloor \cdot \rfloor$ denotes the integer part, 
and we assume $-1/2 \le \R(\kappa) \le 1/2$.

In the context of the generalized Fisher--Hartwig conjecture,
our expansion (\ref{GFH-1}), (\ref{GFH-2}) may be viewed as
a detailed description of the switching between Fisher--Hartwig
branches as a function of the parameter $\kappa$. In physical
terms, this switching between asymptotic branches is a particular
example of a more general notion of ``counting phase transition''
introduced for full-counting-statistics problems in 
Ref.~\onlinecite{2010-IvanovAbanov}.

A practical application of our result is
a convenient method of computing cumulants
of the number of fermions $\hat{Q}$ on a line
segment of length $L$ in the free-fermion problem
described in Section~\ref{sec:formulation}.
While those cumulants may, in principle, be computed 
using the Wick theorem (see, e.g., Ref.~\onlinecite{2011-AIQ}),
the complexity of such calculations grows rapidly with
the order of the cumulant and with the degree of the $1/L$
correction. Remarkably, the same cumulants may be obtained
by the straightforward Taylor expansion of
our series (\ref{GFH-1}), (\ref{GFH-2}) at
$\kappa=0$. Below we list first several
cumulants up to the order $x^{-3}$ obtained
in such a way:
\begin{eqnarray}
	\pi \langle \hat{Q} \rangle &=& x\, ,
 \nonumber \\
	\pi^{2} \langle\langle \hat{Q}^{2}\rangle\rangle &=& 1+\Lambda -\frac{1}{4x^{2}}\cos(2x)
	-\frac{1}{2x^{3}}\sin(2x)+ o(x^{-3})\, ,
 \nonumber \\
	\pi^{3} \langle\langle \hat{Q}^{3}\rangle\rangle &=& \frac{3}{2x} +\frac{3}{2x^{2}}\Lambda \sin(2x)
	-\frac{1}{8x^{3}}+\frac{3}{4x^{3}}(3-4\Lambda)\cos(2x)+ o(x^{-3})\, ,
 \nonumber \\
	\pi^{4} \langle\langle \hat{Q}^{4}\rangle\rangle 
	&=& -\frac{3}{2}\zeta(3)-\frac{15}{4x^{2}} +\frac{6}{x^{2}}\Lambda^{2} \cos(2x)
	-\frac{3}{2x^{3}}(3+4\Lambda)(3-2\Lambda)\sin(2x)+ o(x^{-3})\, ,
 \\
	\pi^{5} \langle\langle \hat{Q}^{5}\rangle\rangle 
	&=& -\frac{5}{2x^{2}}\left[\zeta(3)+8\Lambda^{3}\right]\sin(2x)
	+\frac{165}{8x^{3}} +\frac{5}{x^{3}}\left[\zeta(3)-\Lambda(9+2\Lambda)(9-4\Lambda)\right] \cos(2x)
	+ o(x^{-3})\, ,
 \nonumber \\
	\pi^{6} \langle\langle \hat{Q}^{6}\rangle\rangle 
	&=&  \frac{15}{2}\zeta(5)
	- \frac{30}{x^{2}}\Lambda\left[\zeta(3)+2\Lambda^{2}\right]\cos(2x)
	+\frac{15}{x^{3}}\left[(3-4\Lambda)\zeta(3)-2\Lambda^{2}(3-2\Lambda)^{2}\right] \sin(2x)
	+ o(x^{-3})\, ,
 \nonumber
\end{eqnarray}
where
\begin{equation}
	\Lambda = \log(2x) + \gamma_{E} \, ,
\end{equation}
$x=k_F L$ as before, $\gamma_E$ is the
Euler-Mascheroni constant, and $\zeta(n)$ is the Riemann zeta function
[which arise from the expansion of the Barnes $G$ function
in Eq.~(\ref{C-coefficient})].

\begin{acknowledgments}
We thank V.~Korepin for useful discussions
and N.~Kitanine, K.~Kozlowski, J.~M.~Maillet, N.~Slavnov, and V.~Terras
for comments on the manuscript.
A.G.A.\ was supported by the NSF under the grant DMR-1206790.
\end{acknowledgments}



\begin{thebibliography}{99}
\bibitem{note-degree}
We also observe that the degrees of the polynomials $f_n(\kappa)$ equal
$n+2$, but we could not prove this observation.
\end{thebibliography}
\end{document}